\documentclass[aps,prd,twocolumn,superscriptaddress,nofootinbib]{revtex4-2}

\usepackage[T1]{fontenc}
\usepackage{amsmath,amssymb}
\usepackage{booktabs}
\usepackage{microtype}
\usepackage{siunitx}
\usepackage{xcolor}
\usepackage{hyperref}
\usepackage{pgfplots}
\pgfplotsset{compat=1.18}
\hypersetup{
  colorlinks=true,
  linkcolor=blue!45!black,
  citecolor=green!35!black,
  urlcolor=blue!55!black
}

\definecolor{tracedarkblue}{HTML}{005A8D}
\definecolor{tracelightblue}{HTML}{56B4E9}
\definecolor{adjointdarkorange}{HTML}{C45100}
\definecolor{adjointlightorange}{HTML}{E69F00}
\definecolor{multipletgreen}{HTML}{009E73}

\begin{document}

\title{Exact color sums for multi-gluon amplitudes:\texorpdfstring{\\}{ }
direct, multiplet and symmetric-group Fourier methods}

\author{Rikkert Frederix}
\email{rikkert.frederix@fysik.lu.se}
\affiliation{Department of Physics, Lund University, Box 118, 221 00 Lund, Sweden}
\author{Valentin Hirschi}
\email{hirschi@itp.unibe.ch}
\affiliation{Institute for Theoretical Physics, University of Bern,
Sidlerstrasse 5, 3012 Bern, Switzerland}
\author{Malin Sj\"odahl}
\email{malin.sjodahl@fysik.lu.se}
\affiliation{Department of Physics, Lund University, Box 118, 221 00 Lund, Sweden}

\begin{abstract}
We compare three approaches for computing color-summed tree-level
all-gluon squared matrix elements: evaluation using direct contraction in the
trace and adjoint decompositions, orthonormal multiplet bases, and
fast Fourier transforms (FFT) based on the irreducible
representations of the symmetric group for the trace and adjoint
decompositions.
The multiplet method reduces the color sum to a sum of
absolute squares and utilizes an amplitude recursion labeled by
$SU(3)$ representations. The FFT exploits the relative-permutation
dependence of color overlaps to replace the direct double sum with
smaller independent contractions, using standard color-ordered
partial amplitudes. This invokes the symmetric group at the level of
gluon labels, and therefore makes maximal use of the permutation symmetry.
We compare CPU time and memory use for four
through eleven gluons. At eleven gluons, the direct adjoint
contraction evaluates a fixed-helicity color-summed matrix element in an estimated
\SI{1.7e3}{s} after initialization, compared with \SI{19.8}{s} for the
multiplet recursion and \SI{0.814}{s} for the FFT adjoint method.
The
adjoint FFT is fastest from six through eleven gluons, despite its
factorial scaling compared with the exponential scaling of the
multiplet recursion.
\end{abstract}

\maketitle

\section{Introduction}
\label{sec:intro}

Color-ordered Berends--Giele recursion efficiently computes
tree-level gluon partial amplitudes~\cite{Berends:1987me}. However,
computing an exact color-summed squared matrix element also requires
evaluating all interference terms among the associated color
structures.  At fixed external momenta and helicities, let $\mathcal
M$ denote the full amplitude and let the index $\alpha$ label the
chosen color tensors $C_\alpha$, such that
\begin{equation}
  \mathcal M=\sum_{\alpha} C_{\alpha}\,A_{\alpha},
  \label{eq:color-decomposition}
\end{equation}
where $A_\alpha$ is the corresponding complex kinematic amplitude.
The squared color-summed amplitude is then
\begin{equation}
  \begin{aligned}
    \sum_{\text{colors}}|\mathcal M|^2
    =\sum_{\alpha,\beta}A_{\alpha}^{*}G_{\alpha\beta}A_{\beta},\\
  \end{aligned}
  \label{eq:color-sum}
\end{equation}
where
\begin{equation}
  \begin{aligned}
    G_{\alpha\beta}
    &\equiv\sum_{\text{colors}}C_{\alpha}^{*}C_{\beta}
  \end{aligned}
  \label{eq:color-sumG}
\end{equation}
gives the (real) scalar product matrix between color tensors.

The trace decomposition~\cite{Paton:1969je,Mangano:1987xk} and the adjoint
basis of Del Duca, Dixon and Maltoni~\cite{DelDuca:1999rs} express the
amplitude in terms of color-ordered partial amplitudes.  Their
color tensors are not orthogonal, so the off-diagonal terms in
Eq.~\eqref{eq:color-sumG} must be retained.  In both cases, the number of
orderings grows factorially with the gluon multiplicity, making direct
evaluation of the double sum increasingly expensive.

Multiplet bases offer a different organization of color
space~\cite{Keppeler:2012ih,Sjodahl:2018cca,Sjodahl:2024fqn}.  They
are constructed by coupling gluons through irreducible $SU(3)$
representations, and their color tensors can be chosen
orthonormal. The recursion in multiplet bases entails reordering
gluons into a fixed reference order. The details were worked out in
Ref.~\cite{Bolinder:2025gbj}, where an exponential scaling with the
number of gluons was proven.  Here we present the first numerical
implementation.

The non-orthogonal trace and adjoint decompositions nevertheless have
useful structure, since all gluons enter on equal footing.  Therefore,
in both cases, the overlap between two color tensors depends only on
the relative permutation of their gluon orderings.  A fast Fourier
transform (FFT) on the symmetric group exploits this property to
replace the direct double sum with a set of smaller, independent
matrix contractions.  The Fourier method, which was suggested to us by
\texttt{Codex} running \texttt{ChatGPT 5.6-Sol}, is exact: it retains
every color interference term and does not change the partial
amplitudes.  To our knowledge, this is the first application of a
symmetric-group FFT to exact color summation for QCD matrix elements.

We compare the three approaches of direct trace or adjoint squaring,
multiplet bases and the FFT/symmetric group applied to trace and
adjoint partial amplitudes.  Section~\ref{sec:direct} describes direct
color summation in the trace and adjoint decompositions.
Section~\ref{sec:multiplet} presents the orthogonal multiplet basis
and its recursion.  Section~\ref{sec:fft} explains the symmetric-group
Fourier method for the trace and adjoint color sums. CPU times per
evaluation after initialization and maximum memory use are compared in
section~\ref{sec:results}, and conclusions are drawn in
section~\ref{sec:conclusion}.

\section{Direct color sums in trace and adjoint decompositions}
\label{sec:direct}

\subsection{Trace decomposition}

Tree-level color structures can be written as single traces over gluon color indices.
Cyclic invariance lets us fix one external gluon in the single-trace
decomposition.  For any positive integer $q$, let $S_q$ denote the group of
permutations of $q$ labels; let $\sigma\in S_q$ be one such permutation, and
let $\sigma(i)$ be the image of label $i$.  Permuting the remaining $N-1$ gluons
gives~\cite{Paton:1969je,Mangano:1987xk}
\begin{equation}
  \mathcal M
  =\sum_{\sigma\in S_{N-1}}
   C^{\mathrm{tr}}_{\sigma}
   A\bigl(\sigma(1),\ldots,\sigma(N-1),N\bigr),
  \label{eq:trace-decomposition}
\end{equation}
where $C^{\mathrm{tr}}_\sigma$ is the trace color tensor for ordering $\sigma$,
$A(\sigma(1),\ldots,\sigma(N-1),N)$ is the corresponding partial amplitude,
and there are $P_{\mathrm{tr}}=(N-1)!$ trace orderings, giving $((N-1)!)^2$
terms for the naive color-summed amplitude square.

The trace set is
overcomplete for $SU(3)$ but conceptually simple and convenient because its kinematic coefficients are the usual
color-ordered amplitudes.  Reflection relates each ordering to its reverse,
so the corresponding currents can be shared when evaluating partial
amplitudes.

\subsection{Adjoint decomposition}

The adjoint decomposition places gluons $1$ and $N$ at the ends of a chain of
structure constants and permutes the remaining gluons~\cite{DelDuca:1999rs},
\begin{equation}
  \mathcal M
  =\sum_{\sigma\in S_{N-2}}
   C^{\mathrm{adj}}_{\sigma}
   A\bigl(1,\sigma(2),\ldots,\sigma(N-1),N\bigr),
  \label{eq:adjoint-decomposition}
\end{equation}
where $C^{\mathrm{adj}}_\sigma$ is the chain of structure
constants for ordering $\sigma$,
$A(1,\sigma(2),\ldots,\sigma(N-1),N)$ is the corresponding partial amplitude,
and $P_{\mathrm{adj}}$ is the number of adjoint orderings. 
There are thus $ P_{\mathrm{adj}}=(N-2)!$
partial amplitudes.  The adjoint set is
smaller than the trace set, but its color tensors remain non-orthogonal.
Like the trace set, the adjoint decompositions are only valid for
tree-level processes.

\subsection{Direct contraction}

For either decomposition, let $P$ denote $P_{\mathrm{tr}}$ or
$P_{\mathrm{adj}}$, as appropriate. Let the permutations $\sigma$,
$\tau$, and $\pi$ belong to $S_{N-1}$ or $S_{N-2}$ for the trace and
adjoint decompositions, respectively.  Direct evaluation of
Eq.~\eqref{eq:color-sum} contracts one pair of orderings at a time.
Because the color scalar product matrix is real and symmetric, we
evaluate each diagonal term once, while each off-diagonal pair
contributes twice the real part of its upper-triangular term.  A
common relabeling of both color tensors leaves their overlap
unchanged, so
\begin{equation}
  G_{\sigma\tau}=h(\sigma^{-1}\tau),
  \qquad h(\pi)=h(\pi^{-1}).
  \label{eq:relative-permutation}
\end{equation}
Here $G_{\sigma\tau}$ is an element of the scalar product matrix, Eq.~\eqref{eq:color-sumG}, and thus the overlap of the color tensors for the two orderings. 
The superscript $-1$ denotes permutation inversion, and $h(\pi)$ is the
overlap obtained by placing the first tensor in the reference order and using
$\pi$ as the relative permutation.  Only the $P$ values of $h$ need to be
stored because every row of $G$ is a permutation of this list.  The
contraction still considers $P^2$ pairs, so its time scales as $[(N-1)!]^2$  and $[(N-2)!]^2$ 
for trace and adjoint decompositions, respectively.  These direct contractions provide the
baseline for the comparison below.

For both decompositions, we use Berends--Giele
recursion~\cite{Berends:1987me} to evaluate the ordered amplitudes.  We compute
currents shared by several complete orderings once and reuse them.  Our comparison
therefore includes both partial-amplitude evaluation and final color
contraction, rather than the color contraction alone.

\section{The multiplet basis}
\label{sec:multiplet}

The multiplet bases that we use organize the color space by successively coupling
gluons into irreducible representations of $SU(3)$.  We fix an ordering of the
external gluons, couple the first two to an allowed irreducible
representation, and then couple each subsequent gluon in turn,
requiring the full system to form a singlet. This gives basis vectors
of the form~\cite{Keppeler:2012ih}
\begin{equation}
  \label{eq:MSSKvector}
  \raisebox{-0.5 \height}{\includegraphics[scale=0.4]{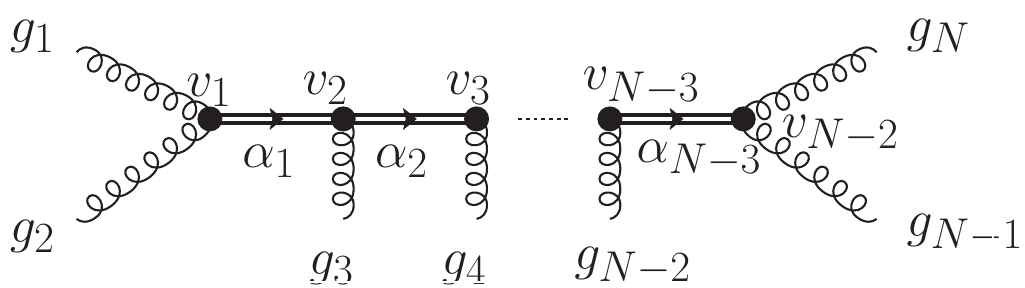}},
\end{equation}
where the double line labeled $\alpha_n$ carries the combined
representation of the first $n+1$ gluons. The vertex multiplicity
labels $v_n$ distinguish independent couplings when a representation
appears more than once in a tensor product, as, for example, for
the two octets in $8\otimes8$. Each basis tensor is specified by its
sequence of intermediate representations and vertex multiplicity
labels. These vectors span the full color space for the external
gluons and can therefore be used at any perturbative order.

With suitable normalization, the basis vectors are orthonormal, i.e.,
\begin{equation}
  \sum_{\text{colors}}
  C_{\alpha}^{*}C_{\beta}=\delta_{\alpha\beta},
  \label{eq:multiplet-orthogonality}
\end{equation}
where now $\alpha$ and $\beta$ collectively denote the representation
sequences and vertex multiplicity labels.
Equation~\eqref{eq:color-sum} therefore becomes
\begin{equation}
\sum_{\text{colors}}|\mathcal M|^2 =\sum_{\alpha}|A_{\alpha}|^2.
  \label{eq:multiplet-sum}
\end{equation}
Thus, once the amplitude coefficients $A_\alpha$ are known, exact color
summation reduces to an ordinary sum of absolute squares.

We compute the amplitude coefficients in this basis using the
off-shell recursion of Ref.~\cite{Bolinder:2025gbj}.  A current is
labeled by a subset of external gluons and by an allowed sequence of
$SU(3)$ representations and vertex multiplicity numbers.  Joining two
currents requires reordering their gluons into the order defining the
larger basis.  We decompose this reordering into exchanges of
neighboring gluons.  Each exchange acts on the representation
sequences through a normalized matrix of Wigner $6j$
coefficients~\cite{Du:2015apa}.

Aside from the representations involved, the Wigner $6j$ coefficients
depend on the normalization, phase, and multiplicity conventions
chosen for the three-point $SU(3)$ vertices. To ensure consistency
with the conventions used for our multiplet bases, we calculate these
coefficients from scratch. For processes with up to $N=11$ gluons, it
suffices to consider all irreducible representations appearing in
$8^{\otimes 5}$. We tabulate the normalized exchange matrices for the
required reorderings using two independent methods: direct contraction
of explicit three-point tensors whose components are Clebsch--Gordan
coefficients, and recursive reduction to the elementary two-line
reorderings described in
Refs.~\cite{Alcock-Zeilinger:2022hrk,Keppeler:2023msu}. Both methods
yield exactly the same coefficient matrices. The complete table
contains 3875 non-zero coefficients and covers every reordering
required for processes with up to eleven gluons.

The representations allowed in repeated adjoint products, while
enforcing that the total $N$ gluons should form a singlet determine
the number of multiplet basis vectors.  For $N=4,\ldots,11$, the
respective counts are $8$, $32$, $145$, $702$, $3598$, $19280$,
$107160$, and $614000$~\cite{Du:2015apa}.  Because the multiplet bases
are valid to all orders, they only become smaller than the trace and
adjoint sets from $N=7$ and $N=12$, respectively.

Orthogonality eliminates the final double sum, whereas the representation-labeled
recursion requires more work than the ordered trace and adjoint
recursions~\cite{Bolinder:2025gbj}.

\section{Symmetric-group FFT for trace and adjoint color sums}
\label{sec:fft}

The trace and adjoint decompositions admit an efficient color
summation because the overlap of two color tensors depends only on the
relative permutation of their gluon orderings. This allows the color
contraction to be expressed through a convolution on the symmetric
group, which a Fourier transform converts into independent matrix
products.

For either decomposition, the partial amplitudes define a function
$A(\sigma)$ on $S_m$, with $m=N-1$ for trace and $m=N-2$ for
adjoint. Using the overlap function $h$ defined in
Eq.~\eqref{eq:relative-permutation}, we write
\begin{align}
(A*h)(\tau)&=\sum_{\sigma\in S_m}A(\sigma)\,h(\sigma^{-1}\tau),\nonumber\\
\sum_{\text{colors}}|\mathcal M|^2&=\sum_{\tau\in S_m}A(\tau)^{*}(A*h)(\tau).\label{eq:conv1}
\end{align}
The standard discrete Fourier transform on a cyclic group converts
convolutions into multiplications of scalar Fourier coefficients.  For
the symmetric group of permutations of the gluon labels, the Fourier
coefficients are instead matrices associated with irreducible
representations, and convolutions become matrix multiplications.
These representations are labeled by partitions $\lambda$ of $m$,
written $\lambda\vdash m$, or equivalently by Young diagrams with $m$
boxes. Denoting the representation dimension by $d_\lambda$, we have
\begin{equation}
\sum_{\lambda\vdash m}d_\lambda^2=m!.
\end{equation}

For a complex-valued function $f$ on $S_m$, define its transformed matrix
$\widehat f_\lambda$ by
\begin{equation}
  \widehat f_\lambda
  =\sum_{\sigma\in S_m}f(\sigma)\rho_\lambda(\sigma),
  \label{eq:group-transform}
\end{equation}
where $\rho_\lambda(\sigma)$ is the real orthogonal
$d_\lambda\times d_\lambda$ matrix representing the permutation $\sigma$ in
the irreducible representation labeled by $\lambda$.
The transform therefore reorganizes the $m!$ partial amplitudes into
one $d_\lambda\times d_\lambda$ matrix for each Young diagram, without
losing information.
Let $\widehat A_\lambda$ be the transform of the partial-amplitude function
$A(\sigma)$, and let $\widehat h_\lambda$ be the transform of the color-overlap
function $h(\sigma)$.  The exact color sum can be written
\begin{equation}
\sum_{\text{colors}}|\mathcal M|^2=\frac{1}{m!}
  \sum_{\lambda\vdash m}d_\lambda
  \operatorname{Tr}\!\left(
    \widehat A_\lambda^\dagger
    \widehat A_\lambda \widehat h_\lambda
  \right),
  \label{eq:fft-color-sum}
\end{equation}
after invoking Parseval's theorem.
Thus, the convolutions in Eq.~\eqref{eq:conv1} become a sum of independent
matrix contractions.
We remark that this block structure follows from the orthogonality relations for
matrix elements of irreducible representations of the symmetric
group, first explored in Ref.~\cite{Zeppenfeld:1988bz}. There is thus a close analogy with multiplet bases, where
representation theory is likewise used to construct an orthogonal
basis. The key distinction is that multiplet bases couple gluons
through irreducible $SU(3)$ representations, whereas here the
irreducible representations are used to transform functions defined
on the symmetric group. Both constructions use Young diagrams
for constructing the irreducible representations, but for the
multiplet bases the symmetrization is done at the level of
fundamental and antifundamental indices~\cite{Keppeler:2012ih}. 

Computing every transformed matrix entry directly from
Eq.~\eqref{eq:group-transform} would still be expensive because it
still scales as a squared factorial.  Fast transforms on finite groups instead
reuse partial results~\cite{DiaconisRockmore1990}.  For the symmetric
group, we evaluate the transform recursively along the subgroup chain
\begin{equation}
  S_1\subset S_2\subset\cdots\subset S_m,
  \label{eq:subgroup-chain}
\end{equation}
where $S_i$ is the permutation group on $i$ labels for $i=1,\ldots,m$.
For $i=1,\ldots,m-1$, the inclusion $S_i\subset S_{i+1}$ leaves label
$i+1$ unchanged.  Partial results at one level are reused at the next.
Along this chain, the matrices representing exchanges of adjacent
gluons split into one- and two-dimensional blocks, reducing the work
at each step: the overal scaling goes down from factorially squared to
factorial times a polynomial~\cite{ClausenBaum1993,Maslen1998}.

For each multiplicity $N$, we compute the $\widehat h_\lambda$
matrices once and reuse them for all phase-space points and helicity
assignments. Together, these matrices contain $\sum_{\lambda\vdash
  m}d_\lambda^2=m!$ entries.  Each matrix-element evaluation then
requires transforming the partial amplitudes and performing the
contractions in Eq.~\eqref{eq:fft-color-sum}.  The FFT does not change
the partial amplitudes and their number; rather, it reorganizes the
color contraction into independent blocks and avoids the direct sum
over $P^2$ pairs of orderings.

\section{Numerical comparison}
\label{sec:results}

\begin{table*}[t]
  \caption{CPU time for one evaluation of a fixed-helicity, colour-summed
  squared matrix element after initialization.  Initialization and the first
  evaluation are excluded.  Direct denotes colour-overlap-matrix contraction
  without a Fourier transform, in either the trace decomposition or the
  adjoint basis.  FFT denotes the exact symmetric-group Fourier method.
  Asterisks mark estimates, not measurements.}
  \label{tab:evaluation-times}
  \centering
  \setlength{\tabcolsep}{5.5pt}
  \begin{tabular}{@{}c rr r rr@{}}
  \toprule
  $N$ & \multicolumn{2}{c}{direct colour sum} &
  \multicolumn{1}{c}{orthogonal basis} &
  \multicolumn{2}{c}{Fourier colour sum} \\
  \cmidrule(lr){2-3}\cmidrule(lr){4-4}\cmidrule(lr){5-6}
  & trace & adjoint & multiplet & trace FFT & adjoint FFT \\
  \midrule
  4  & \SI{0.729}{\micro\second} & \SI{0.543}{\micro\second} & \SI{0.923}{\micro\second} & \SI{0.968}{\micro\second} & \SI{0.645}{\micro\second} \\
  5  & \SI{6.23}{\micro\second}  & \SI{2.44}{\micro\second}  & \SI{6.38}{\micro\second}  & \SI{4.41}{\micro\second}  & \SI{2.62}{\micro\second} \\
  6  & \SI{138}{\micro\second}   & \SI{15.3}{\micro\second}  & \SI{62.3}{\micro\second}  & \SI{26.2}{\micro\second}  & \SI{13.3}{\micro\second} \\
  7  & \SI{6.16}{\milli\second}  & \SI{0.196}{\milli\second} & \SI{0.718}{\milli\second} & \SI{0.186}{\milli\second} & \SI{0.0832}{\milli\second} \\
  8  & \SI{0.381}{\second}       & \SI{5.78}{\milli\second}  & \SI{8.26}{\milli\second}  & \SI{1.62}{\milli\second}  & \SI{0.620}{\milli\second} \\
  9  & \SI{29.7}{\second}        & \SI{0.303}{\second}       & \SI{0.104}{\second}       & \SI{0.0155}{\second}      & \SI{0.00523}{\second} \\
  10 & \SI{2.4e3}{\second}\textsuperscript{*} & \SI{20.9}{\second} & \SI{1.49}{\second} & \SI{0.215}{\second} & \SI{0.0589}{\second} \\
  11 & \SI{2.4e5}{\second}\textsuperscript{*} & \SI{1.7e3}{\second}\textsuperscript{*} & \SI{19.8}{\second} & \SI{3.08}{\second} & \SI{0.814}{\second} \\
  \bottomrule
  \end{tabular}

  \vspace{0.5ex}
  \begin{minipage}{0.97\textwidth}
  \footnotesize
  \textsuperscript{*}Estimated from the last measured direct result, assuming
  that the time is proportional to the square of the number of
  colour orders, with $P_{\mathrm{tr}}=(N-1)!$ and
  $P_{\mathrm{adj}}=(N-2)!$.
  \end{minipage}
\end{table*}

\subsection{Setup and checks}

We study $gg\to(N-2)g$, summing the colors of all incoming and
outgoing gluons while holding their helicities fixed.  For each
$N=4,\ldots,11$, we apply all five methods to the same phase-space
point and the same helicity assignment, chosen to give a nonzero
matrix element. We use the general off-shell recursion for all
helicity configurations, even when an analytic expression is
available, so special-case formulas do not affect the comparison. The
timing is therefore independent of the chosen nonzero helicity
configuration.

We compiled the five implementations with \texttt{GNU Fortran 13.3} at
optimization level \texttt{-O3} and ran each on a single core of an
\texttt{Intel Core i7-8700K}. The timing excludes initialization and the
first evaluation, during which any remaining $N$-dependent data are
prepared.  The reported CPU time per subsequent matrix-element
evaluation is the median of ten measurements.

For each method, we measured maximum physical memory in a separate process;
the values include initialization and all evaluations.

For every tested configuration, all completed calculations agree on
the color-summed squared matrix element.  The largest pairwise
relative difference between nonzero results is $8.1\times10^{-12}$, at
ten gluons.  Extrapolated times are excluded from these checks.

\subsection{Evaluation time after initialization}

\begin{figure}[t]
  \centering
  \begin{tikzpicture}
    \begin{axis}[
      width=0.97\columnwidth,
      height=0.70\columnwidth,
      xlabel={number of external gluons $N$},
      ylabel={CPU time per evaluation [s]},
      xmin=3.7,xmax=11.3,
      xtick={4,5,6,7,8,9,10,11},
      ymode=log,
      ymin=1e-7,ymax=1e6,
      ymajorgrids=true,
      grid style={draw=black!12,line width=0.35pt},
      axis line style={draw=black!65,line width=0.5pt},
      tick align=outside,
      legend style={at={(0.02,0.98)},anchor=north west,
                    draw=black!18,fill=white,fill opacity=0.94,
                    text opacity=1,font=\scriptsize,rounded corners=1pt,
                    legend columns=2,
                    /tikz/every even column/.append style={column sep=7pt}},
      tick label style={font=\scriptsize},
      label style={font=\small},
      unbounded coords=jump,
    ]
      \addplot+[mark=square*,mark size=2.3pt,thick,densely dashed,tracedarkblue,
                mark options={solid,fill=tracedarkblue,draw=tracedarkblue}]
        table[x=N,y=trace_direct] {data/colour_sum_times.dat};
      \addlegendentry{trace, direct}
      \addplot+[mark=diamond*,mark size=2.5pt,thick,tracelightblue,
                mark options={solid,fill=tracelightblue,draw=tracelightblue}]
        table[x=N,y=trace_fft] {data/colour_sum_times.dat};
      \addlegendentry{trace, FFT}
      \addplot+[mark=triangle*,mark size=2.6pt,thick,densely dashed,adjointdarkorange,
                mark options={solid,fill=adjointdarkorange,draw=adjointdarkorange}]
        table[x=N,y=adjoint_direct] {data/colour_sum_times.dat};
      \addlegendentry{adjoint, direct}
      \addplot+[mark=pentagon*,mark size=2.4pt,thick,adjointlightorange,
                mark options={solid,fill=adjointlightorange,draw=adjointlightorange}]
        table[x=N,y=adjoint_fft] {data/colour_sum_times.dat};
      \addlegendentry{adjoint, FFT}
      \addplot+[mark=*,mark size=2.2pt,thick,multipletgreen,
                mark options={solid,fill=multipletgreen,draw=multipletgreen}]
        table[x=N,y=multiplet] {data/colour_sum_times.dat};
      \addlegendentry{multiplet}
      \addplot+[no marks,thick,densely dotted,tracedarkblue,forget plot]
        table[x=N,y=trace_extension] {data/colour_sum_times.dat};
      \addplot+[no marks,thick,densely dotted,adjointdarkorange,forget plot]
        table[x=N,y=adjoint_extension] {data/colour_sum_times.dat};
      \addplot+[only marks,mark=square,mark size=2.7pt,
                mark options={fill=white,draw=tracedarkblue,line width=0.8pt},forget plot]
        table[x=N,y=trace_estimate] {data/colour_sum_times.dat};
      \addplot+[only marks,mark=triangle,mark size=3.0pt,
                mark options={fill=white,draw=adjointdarkorange,line width=0.8pt},forget plot]
        table[x=N,y=adjoint_estimate] {data/colour_sum_times.dat};
    \end{axis}
  \end{tikzpicture}
  \caption{CPU time to evaluate one fixed-helicity, color-summed squared
  matrix element after initialization.  Dashed lines show direct trace and
  adjoint color sums; solid blue and orange lines show their exact Fourier
  counterparts; the solid green line shows the multiplet method.  Dotted
  extensions and open symbols indicate the three estimates identified by
  asterisks in Table~\ref{tab:evaluation-times}, while filled symbols indicate
  measurements.}
  \label{fig:evaluation-times}
\end{figure}
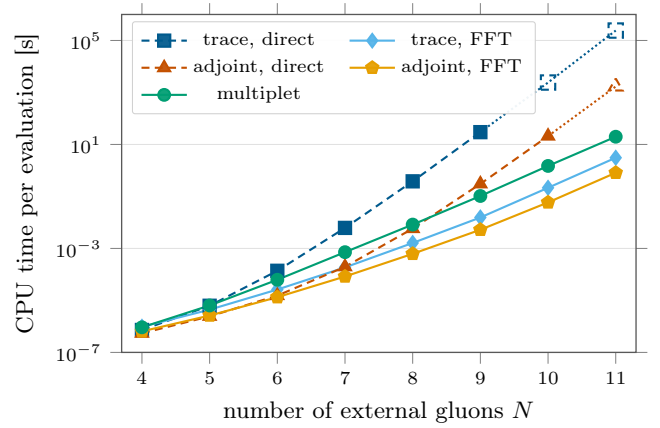

Table~\ref{tab:evaluation-times} and Fig.~\ref{fig:evaluation-times} summarize
the timing results.  Direct adjoint contraction is fastest at four and five
gluons; adjoint with the Fourier transform is fastest from six through eleven
gluons.  The direct double sum becomes costly at around eight gluons: direct trace
takes \SI{0.381}{s}, whereas trace with the transform takes \SI{1.62}{ms}.
At nine gluons, the corresponding times are \SI{29.7}{s} and \SI{15.5}{ms},
a factor of about $1900$.

The multiplet method avoids the direct double sum.  It is faster than
direct trace at six gluons and comparable to direct adjoint at eight
gluons. At eleven gluons, the multiplet evaluation takes \SI{19.8}{s},
versus an estimated \SI{2.41e5}{s} for direct trace and \SI{1.69e3}{s}
for direct adjoint, and measured times of \SI{3.08}{s} and
\SI{0.814}{s} for trace and adjoint with the transform,
respectively. The adjoint Fourier method is therefore fastest at this
multiplicity.

\subsection{Memory use}

The following table gives maximum memory use at the four largest
multiplicities.  A dash indicates that the corresponding run did not finish.
\begin{center}
  \small
  \setlength{\tabcolsep}{3.2pt}
  \begin{tabular}{@{}l rrrr@{}}
  \toprule
  method & $N=8$ & $N=9$ & $N=10$ & $N=11$ \\
  \midrule
  trace, direct & \SI{4.24}{MiB} & \SI{14.0}{MiB} & --- & --- \\
  adjoint, direct & \SI{3.34}{MiB} & \SI{7.66}{MiB} & \SI{43.9}{MiB} & --- \\
  multiplet     & \SI{10.1}{MiB} & \SI{55.4}{MiB} & \SI{425}{MiB} & \SI{3.96}{GiB} \\
  trace, FFT    & \SI{4.43}{MiB} & \SI{14.8}{MiB} & \SI{111}{MiB} & \SI{1.09}{GiB} \\
  adjoint, FFT  & \SI{3.54}{MiB} & \SI{7.86}{MiB} & \SI{44.6}{MiB} & \SI{398}{MiB} \\
  \bottomrule
  \end{tabular}
\end{center}

At every multiplicity with a direct result, the direct and Fourier variants
use similar amounts of memory, since the FFT recursion requires
relatively little additional storage.
Our direct implementations store only the $P$ distinct overlaps from
Eq.~\eqref{eq:relative-permutation}, not the full $P\times P$ matrix.  Within
each decomposition, the direct and Fourier methods therefore differ far more
in evaluation time than in stored color data.

At eleven gluons, memory use is \SI{3.96}{GiB} for the multiplet method,
\SI{1.09}{GiB} for trace with the transform, and \SI{398}{MiB} for adjoint
with the transform.  For each of the three methods with results at both
multiplicities, memory use grows by about an order of magnitude from ten to
eleven gluons, suggesting that memory may become an important limitation
beyond the range studied here.

\section{Conclusion and outlook}
\label{sec:conclusion}

We have compared three exact approaches for evaluating fixed-helicity,
color-summed squared matrix elements for tree-level all-gluon
processes: direct evaluation using the trace and adjoint
decompositions, orthogonal multiplet bases, and a fast
Fourier transform based on the irreducible representations of the
symmetric group
for the trace and adjoint decompositions.
We find that both the multiplet bases, and the fast Fourier
transform significantly speed up the color-summed amplitude computations
at high multiplicities.

The multiplet bases also uses symmetric-group Young tableaux to
organize irreducible $SU(3)$ representations, but the associated
permutations act on the fundamental and antifundamental color indices
used to construct the gluon color tensors. Orthogonality reduces the
final color sum to a
sum of absolute squares, but the amplitude recursion is more
involved for the multiplet bases. In total, the multiplet bases substantially
outperforms the direct contractions at high multiplicity. However, for the all-gluon processes studied here,
the symmetric-group Fourier method is faster still. It retains the trace or adjoint
partial amplitudes while exploiting the relative permutation
dependence of their color overlaps to replace the direct double sum
with exact contractions labeled by Young diagrams.

At eleven gluons, the CPU times per evaluation of the matrix elements after initialization
are \SI{3.08}{s} for the trace decomposition with the transform and \SI{0.814}{s} for the
adjoint decomposition with the transform, compared to \SI{19.8}{s} using the multiplet
basis, and estimates of \SI{2.4e5}{s} and \SI{1.7e3}{s}, respectively,
for the direct trace and adjoint decompositions. The computation time in
multiplet bases scales as an exponential with the number of gluons,
whereas the FFT method scales as a single factorial. Despite this, the
FFT with adjoint bases are fastest from six through eleven gluons. Fourier
(and direct) methods also use less memory at this multiplicity,
despite the multiplet bases's simpler final color sums.

Processes with external quarks are a natural next application.  The
trace decomposition can be straightforwardly extended to include
quarks.  The pure-gluon adjoint chain cannot be carried over unchanged
because quarks and antiquarks carry fundamental and antifundamental
color.  For one quark--antiquark pair, the corresponding color
decomposition uses an open chain of fundamental generators, with the
gluons permuted between the quark and antiquark.  For several pairs,
the established generalization also tracks the placement and nesting
of the quark lines~\cite{Johansson:2015oia,Melia:2015ika}.  Multiplet
bases can accommodate external quarks, antiquarks, gluons, higher
representations~\cite{Sjodahl:2024fqn} and trivially higher orders,
while remaining orthogonal. The basis construction from
Ref.~\cite{Sjodahl:2018cca} allows for a recursion which parallels the
gluon recursion implemented here.

The FFT method with the symmetric group should be applicable for
processes with mixed particle content, albeit to a lower extent,
since the symmetric group can only be applied to particles of the same kind.
For one quark--antiquark pair, the open-chain color overlaps again depend only on the
relative gluon ordering, so the same symmetric-group transform applies
once the required open-chain overlaps have been computed.  With
several pairs, one could apply the transform to gluon permutations
within each quark-line arrangement, but different arrangements remain
coupled. To what extent such partial transforms reduce the total cost
remains to be determined.

The implementation used for the numerical results in this paper is
publicly available at
\url{https://github.com/rikkert-frederix/AllGluonsMultipletFFT}.\\

\textit{We would like to end this paper with a personal note. We find
  it exciting that the symmetric-group FFT method for trace and
  adjoint color sums was suggested to us by an AI model in response to
  our request to ``optimize the implementation.'' To us, this is a sign
  that AI has become more than a tool for accelerating familiar tasks:
  it can bring ideas from distinct fields into our own, opening new
  avenues for progress.}

\begin{acknowledgments}
V.H.~acknowledges support from the Swiss National Science Foundation
under grant PCEFP2\_203335.
\end{acknowledgments}

\bibliographystyle{apsrev4-2}
\bibliography{references}

\end{document}